# Noise Estimation and Suppression in Quantitative EMCD Measurements

Hitoshi Makino[a,*], Bernd Rellinghaus[a], Sebastian Schneider[a], Axel Lubk[b,c], and Darius Pohl[a,*]

[a] *Dresden Center of Nanoanalysis, Technische Universität Dresden, 01069 Dresden, Germany*

[b] *Leibniz Institute for Solid State and Materials Research Dresden, 01069 Dresden, Germany*

[c] *Institute of Solid State and Materials Physics, Technische Universität Dresden, 01069 Dresden, Germany*

\* Corresponding author at: Dresden Center of Nanoanalysis, Technische Universität Dresden, 01069 Dresden, Germany

Tel.: +49351463410851.

*E-mail address:* Hitoshi.makino@tu-dresden.de (H. Makino)

Darius.pohl@tu-dresden.de (D. Pohl)

## Abstract

Quantitative electron magnetic circular dichroism (EMCD) in transmission electron microscopy (TEM) enables the measurement of magnetic moments with elemental and atomic site sensitivity, but its practical application is fundamentally limited by noise. This study presents a comprehensive methodology for noise estimation and suppression in EMCD measurements, demonstrated on Ti-doped barium hexaferrite lamellae. By employing a classical three-beam geometry and long-term acquisition of electron energy-loss spectra, we systematically analyze the signal-to-noise ratio (SNR) across individual energy channels using bootstrap statistics. A robust energy alignment procedure based on the neighboring Ba-$M_{4,5}$ edges with an adequate energy upsampling is introduced to minimize systematic errors from energy misalignment. The impact of detector noise, particularly from CMOS-based EELS cameras, is evaluated through variance-to-mean analysis and described by the noise amplification coefficients, revealing that detector-amplified shot noise is the dominant noise source. We recommend a stricter SNR threshold for reliable EMCD detection and quantification, ensuring that critical spectral features such as the Fe-$L_{2,3}$ peaks meet the requirements for quantitative analysis. The approach also provides a framework for determining the minimum electron dose necessary for valid measurements and can be generalized to scintillator-based or direct electron detectors. This work advances the reliability of EMCD as a quantitative tool for magnetic characterization at the nanoscale with unknown magnetic structures. The proposed procedures lay the groundwork for improved error handling and SNR optimization in future EMCD studies.

# 1. Introduction

Electron energy-loss magnetic chiral dichroism (EMCD)—the electron analogue of the well-established X-ray magnetic circular dichroism (XMCD)—has emerged as a transformative technique within the realm of transmission electron microscopy (TEM). Based on electron energy loss spectroscopy (EELS), EMCD enables the measurement of magnetic moments with elemental, atomic site, and oxidation state sensitivity in ferro- and ferrimagnetic materials. Unlike XMCD, which relies on circularly polarized X-rays to excite core electrons with orbital angular momentum selectivity, EMCD exploits orbital angular momentum selective inelastic scattering of electrons to achieve magnetic sensitivity[1–3].

As illustrated schematically in figure 1, in classical three-beam EMCD, inelastic scattering of electrons from an initial state comprising of both direct and diffracted beams results in final states generated from superpositions of transitions involving particular initial beams. The amplitude of such transitions is described by the mixed dynamic form factor (MDFF) [4]. If, due to dynamic diffraction, the diffracted beams exhibit a $\pi/2$ phase shift with respect to the direct beam, the imaginary part of the MDFF containing the EMCD signal is maximized on a Thales circle defined by the relative scattering vectors (black break line circles in figure 1). In the positions marked by the red and blue arrows in figure 1, the momentum transfer of the electron wave is circularly polarized, which ultimately allows to extract magnetic information similar to the XMCD setup from these spectra [1,3].

Since its initial demonstration in 2003 [1], EMCD has undergone significant methodological developments, particularly regarding the improvement of the signal-to-noise ratio (SNR) and the reliable extraction of weak magnetic signals at nanometer spatial resolution [3,5–9]. Despite these advances, noise remains a fundamental limitation in the quantitative interpretation and practical application of EMCD. Understanding the noise behavior in EMCD signals is thus crucial for both optimizing experimental protocols and advancing the technique's sensitivity and spatial resolution.

This work systematically investigates the noise characteristics inherent to EMCD measurements and aims at establishing a stable measurement procedure. By disentangling the noise sources and analyzing their impact on the EMCD signal, we aim to provide a comprehensive framework for improving SNR and enabling robust quantitative analyses of the magnetization at the atomic scale.

Several experimental factors contribute to the EMCD signal. Dynamic elastic scattering, which depends on the sample's crystal structure, orientation and thickness, leads to a non-trivial thickness dependence of EMCD signal [10,11]. Plural inelastic scattering leads to a convolution of the EMCD core-loss spectra with the low-loss spectrum (e.g., containing plasmons). Sample mistilt causes asymmetric diffraction patterns, further complicating the extraction of the EMCD signal [12,13]. Additionally, aberrations and limitations of the EELS detector, such as energy-correlated noise specific to a complementary metal-oxide-semiconductor (CMOS) based cameras, can introduce energy-correlated fixed pattern.

A comprehensive understanding of these noise sources is essential for advancing EMCD as a quantitative technique, and recent studies have begun to address these challenges. For example, Thersleff et al. systematically analyzed the SNR behavior of Gaussian-fitted EMCD signals in bcc iron, providing valuable insights into the statistical nature of noise in processed EMCD spectra [14]. Hasan et al. further investigated the statistical distribution of the ratio of magnetic spin to orbital moments—so called $m_L/m_S$ ratio—calculated from EMCD sum rules. They showed that there is a noise-dependent bias of the EMCD signal and specifically, that higher noise levels tend to artificially inflate the extracted magnetic moment values [15]. However, these studies focus primarily on processed or integrated signals, leaving a gap in our understanding of how noise in each energy channel of raw EEL spectra propagates to the EMCD signal and quantities derived from that.

To address these gaps, we have conducted long-term EMCD measurements to accumulate sufficient data for a robust statistical analysis of SNR across individual energy channels. By selecting a beam-stable oxide sample and employing the classical three-beam geometry over a macroscopic region, we minimized the effects of beam-induced damage and sample instability. Furthermore, we developed and implemented an EMCD signal extraction methodology specifically designed to avoid gain variation associated with the CMOS-based EELS detector [16], thereby improving the reliability of our noise analysis.

Through this approach, our study aims to provide a detailed characterization of noise in EMCD signals, offering new insights into its origins, statistical properties, and impact on quantitative magnetic measurements. This work thus lays the foundation for improved error handling and SNR optimization strategies, which are critical for the continued development and application of EMCD at the nano and atomic scale.

For this study, we selected single-crystal lamellae of $BaFe_{11}TiO_{19}$ (M-type barium hexaferrite, BH [17]) as the model system. This material offers an advantage for high-precision EMCD investigations since the Ba-$M_{4,5}$ edge, which is at bit higher energy than the Fe-$L_{2,3}$ edge. This Ba-$M_{4,5}$ spectral proximity to Fe-$L_{2,3}$ edges enable more accurate and sensitive correction of energy shifts between spectra of opposite chirality, which proves a critical factor for minimizing systematic errors in EMCD signal extraction.

M-type barium hexaferrite is also of significant scientific interest due to its tunable magnetic properties. The coercivity and maximum magnetization can be precisely controlled by substituting $Fe^{3+}$ ions with non-magnetic cations of different valence, such as $Sr^{2+}$ or $Ti^{4+}$. Recent studies have demonstrated that Ti doping enhances the uniaxial magnetic anisotropy [18]. The introduction of $Ti^{4+}$ ions lead to a local charge compensation, resulting in the reduction of neighboring $Fe^{3+}$ to $Fe^{2+}$. This mixed-valence state provides a unique opportunity to study the decomposition of the EMCD signal into contributions from $Fe^{3+}$ and $Fe^{2+}$ [19] under the assumption that the total EMCD response can be described as a linear combination of their respective EMCD signals.

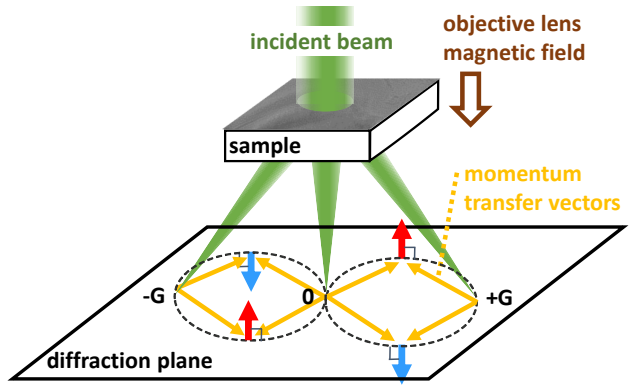

**Figure** 1: Schematic illustration of the EMCD measurement principle in a classical three-beam setup. The sample is magnetized by the objective lens magnetic field of 2 T. Yellow arrows indicate the momentum transfer vectors during chiral inelastic scattering events, defined as the difference between the inelastic (final) and elastic (initial) scattering vectors. Red and blue arrows highlight regions in reciprocal space corresponding to different chiral components of the EMCD signal, respectively, which are essential for isolating the dichroic magnetic signal.

## 2. Experimental

### 2.1 Sample description

We investigated 20 nm thin single-crystalline lamellae of Ti-doped barium hexaferrite (BaFe$_{11}$TiO$_{19}$). Details about crystal growth can be found in [18]. The strong magnetic field of ~2 T generated by the objective lens along the optical axis effectively eliminated magnetic domains within this specimen (magnetic coercive field at room temperature, $H_c$ = 0.2~0.5 T [20]), thereby ensuring a uniform magnetic state during the measurements. In its undoped system, the iron sites in barium hexaferrite predominantly exhibit the Fe$^{3+}$ oxidation state. Upon partial substitution of Fe with Ti$^{4+}$ at 4f$_2$-sites [21], the neighboring Fe$^{3+}$ ions are reduced to Fe$^{2+}$ to maintain charge neutrality, resulting in a mixed-valence state within the lattice. This change in oxidation state leads to a modification of the spin states of the Fe atoms [22].

Undoped Barium hexaferrite furthermore exhibits a high uniaxial magnetocrystalline anisotropy constant ($K_c$ = 3.3 × 10$^5$ J m$^{-3}$) along the [001] crystallographic direction. Since EMCD is sensitive to moments parallel to the electron beam only, the specimen was cut perpendicular to the [001] axis using focused ion beam (FIB) milling, then thinned to approximately 20 nm by Argon-milling, and mounted onto a copper grid for TEM observation. This careful sample preparation ensures both structural integrity and optimal magnetic alignment for high-resolution EMCD analysis.

**3-beam diffraction condition**

The measurements were conducted using a JEOL JEM-F200C transmission electron microscope, operated at an acceleration voltage of 200 kV. The three-beam diffraction condition was adjusted, since this allows for the compensation of asymmetric nonmagnetic contributions to the EMCD signal [23]. All experiments were performed in diffraction mode, employing a beam with a semi-convergence angle of approximately 0.1 mrad and a collection semi-angle of the EELS spectrometer of 4.2 mrad. The corresponding diffraction pattern and intensity line profiles of the three-beam geometry are shown in figure 2.

Careful selection of the diffraction condition is crucial to adjust and maintain the diffraction symmetry in the EMCD experiment; even slight deviations can introduce asymmetries that compromise the accuracy of the magnetic signal extraction [13]. During our measurements, we ensured that the difference in the intensities of the two primary ([110] and [$\bar{1}\bar{1}$0]) diffraction spots stay below 10%, thereby preserving the required orientation and symmetry for reliable EMCD analysis.

As illustrated in Figure 2 (a), the EELS entrance aperture was placed on one of the chiral positions in order to acquire time-series of chiral spectra one by one. Here we defined momentum coordinate axes $k_x$ and $k_y$ along [110] and [$\bar{1}$10] directions respectively. According to the quadrant in this coordinate system, EELS entrance aperture positions were named (+ +), (− +), (− −) and (+ −). The chirality of the spectra at the indicated positions can be defined based on the cross product of the momentum transfer vectors (i.e., the circular polarization of the exchanged virtual photon) and the direction of the magnetization of the sample. Specifically, we denoted the spectra as "red" (+ +), (− −) or "blue" (+ −), (− +) chirality for instance, depending on the relative orientation of these vectors. This rigorous approach to diffraction geometry and chirality assignment is essential for the accurate extraction and interpretation of the EMCD signal.

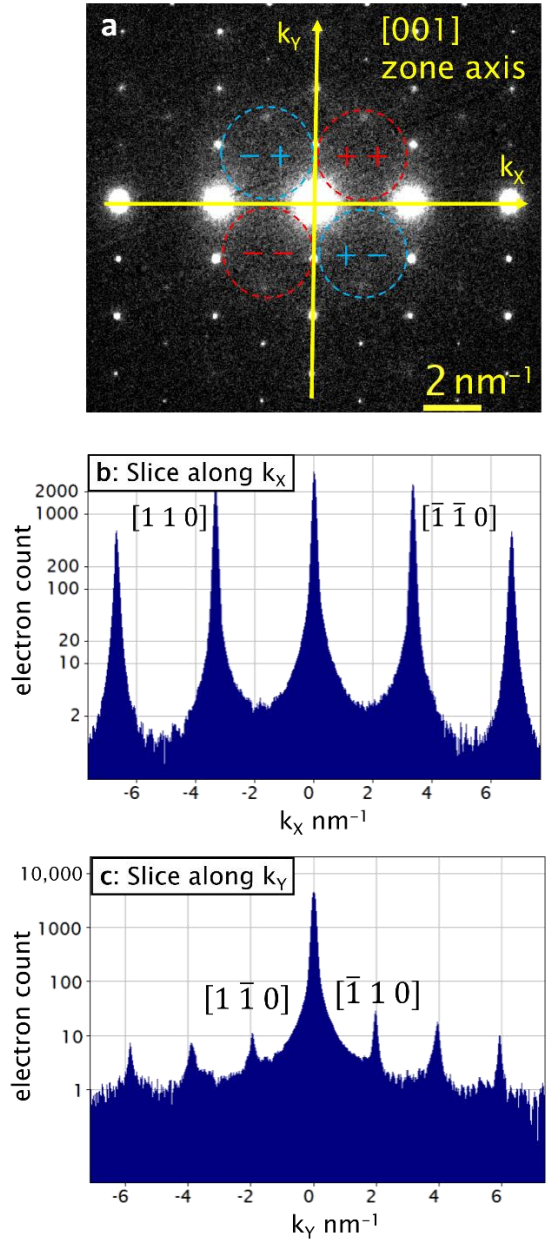

**Figure** 2: Experimental diffraction conditions of the EMCD experiment. **(a)** diffraction pattern of the sample oriented in three-beam condition. Yellow lines ($k_x$ and $k_y$) represent the coordinate system and the diffraction patterns intensity profiles along $k_x$ and $k_y$ axis are shown in **(b)** and **(c)**. The positions of the EELS detector entrance aperture are shown as red and blue circles with the color indicating the different chiralities.

## 2.2 EELS acquisition

For EMCD signal acquisition, we used a Gatan Continuum® Dual EELS/EFTEM system (i.e. GIF camera), which employs a CMOS-based detector. A total of 2,000 frames (spectra) were acquired for a single dataset in dual EELS mode, with a dispersion of 0.15 eV per channel. Employing an exposure time of 1.0 s per frame, the maximal intensity of the zero-loss peak (ZLP) per frame ranged from approximately 40 to 50 electrons. The

Gatan Digital Micrograph® software provides a gain averaging function, a compensation for dark current and gain variation (fixed pattern) through its "HQ dark reference" [24,25]. By default, the number of dark current acquisition frames is set to $3\sqrt{N}$, where $N$ is the total number of the EELS acquisition frames. However, Haruta *et al.* have demonstrated that this default setting is statically insufficient for accurate dark current compensation [26]. Based on their findings, we averaged 1,000 dark current frames—half the number of experimental frames—which was considered sufficient to provide a reliable dark current reference for our measurements. Bosman *et al.* have proposed improved gain averaging by wobbling the energy position of the spectra using the drift tube of the spectrometer during the acquisition (so called *binned gain averaging* method [24]). This method can significantly suppress the remaining gain variation, due to the beam blanking lag in STEM mode with short acquisition times as well as the high dynamic range of the signal. We therefore implemented binned gain averaging with a width of 150 energy channels, corresponding to an energy range of 22.5 eV, during EELS acquisition.

## 2.3 EMCD signal extraction

**Alignment of the energy axes between EEL spectra acquired for different chirality**

Due to the sharpness of the Fe-$L_3$ edges, the EMCD signal is very sensitive to energy misalignment. To mitigate this issue, we implemented a sub-channel energy alignment procedure utilizing the Ba-$M_{4,5}$ edges. To interpolate the spectrum, which is required to detect and correct for sub-channel shifts, we adopted a one-dimensional image interpolation method employing the Lanczos kernel, also referred to as a product of sinc kernels with boundary condition. The Lanczos3 kernel $Lanc(x)$, defined as [27,28]

$$Lanc(x) = \begin{cases} \mathrm{sinc}(x)\mathrm{sinc}(x/3) & |x| \leq 3 \\ 0 & \text{otherwise} \end{cases} \tag{1}$$

is convolved with the original spectra in order to obtain interpolated values at arbitrary energies in between the original channels. The advantages of utilizing the Lanczos3 kernel over simpler methods, such as nearest neighbor or bilinear interpolation, include its superior band-pass characteristics, which effectively minimize blocky artifacts and blurring.

Following the upsampling of the spectra by a factor of 64 employing the Lanczos3 kernel, the energy shifts are corrected in two consecutive steps. First, random energy offsets within one chiral dataset are corrected based on the Fe-$L_{2,3}$ edges because the Ba-$M_{4,5}$ edges in a single spectrum are often too weak for evaluating the correlation. Subsequently, we averaged over the 2,000 EEL spectra of the same chiral type. Second, the energy shifts of $(++)$, $(--)$, $(+-)$, and $(-+)$ averaged chiral spectra are adjusted using the Ba-$M_{4,5}$ edges as non-magnetic standards. All shifts are computed from the correlation function between one of the spectra (in our case, the averaged $(++)$ spectrum) and the selected spectrum.

This systematic approach ensures that the EMCD measurements are both accurate and reliable, facilitating the extraction of meaningful magnetic information from the spectra.

**Background subtraction and normalization**

Each spectrum from different EELS apertures exhibits slight deviations in the background, primarily attributed to zone-axis misalignment of the crystal. To address this discrepancy, we applied a linear fitting approach consisting of normalizing the averaged "blue" chiral spectrum by the averaged "red" chiral spectrum. First, we divided the summed "red" chiral spectrum by the summed "blue" chiral spectrum along the energy axis. Secondly, we applied linear fit with fitting windows of a pre-edge (680~700 eV) and a post-edge (730~ 750 eV), avoiding Fe-$L_{2,3}$ edges themselves. The fitted result is shown in Figure 3 (b). Note that this fit can lose its accuracy when the two spectrum have too small intensities. Therefore, we must carefully check whether the fit is stable. When the chiral spectra have too small intensity, the conventional post Fe-$L_3$ edge normalization would be adequate. Next, the original "blue" chiral spectrum is multiplied by the slope obtained by the fit, and then the total EMCD signal is extracted by subtracting the normalized "blue" from the "red" chiral spectrum as shown in the following Equation.

$$s_{\text{EMCD}} = s_{\text{red}} - s_{\text{blue}} \tag{2}$$

Here, $s_{\text{red}} = s_{++} + s_{--}$ denotes averaged "red" spectrum over the 2,000 frames and $s_{\text{blue}} = s_{+-} + s_{-+}$ the average "blue" spectrum over the 2,000 frames normalized by linear fitting, respectively. This method effectively corrects the background variations, enhancing the accuracy of the spectral analysis.

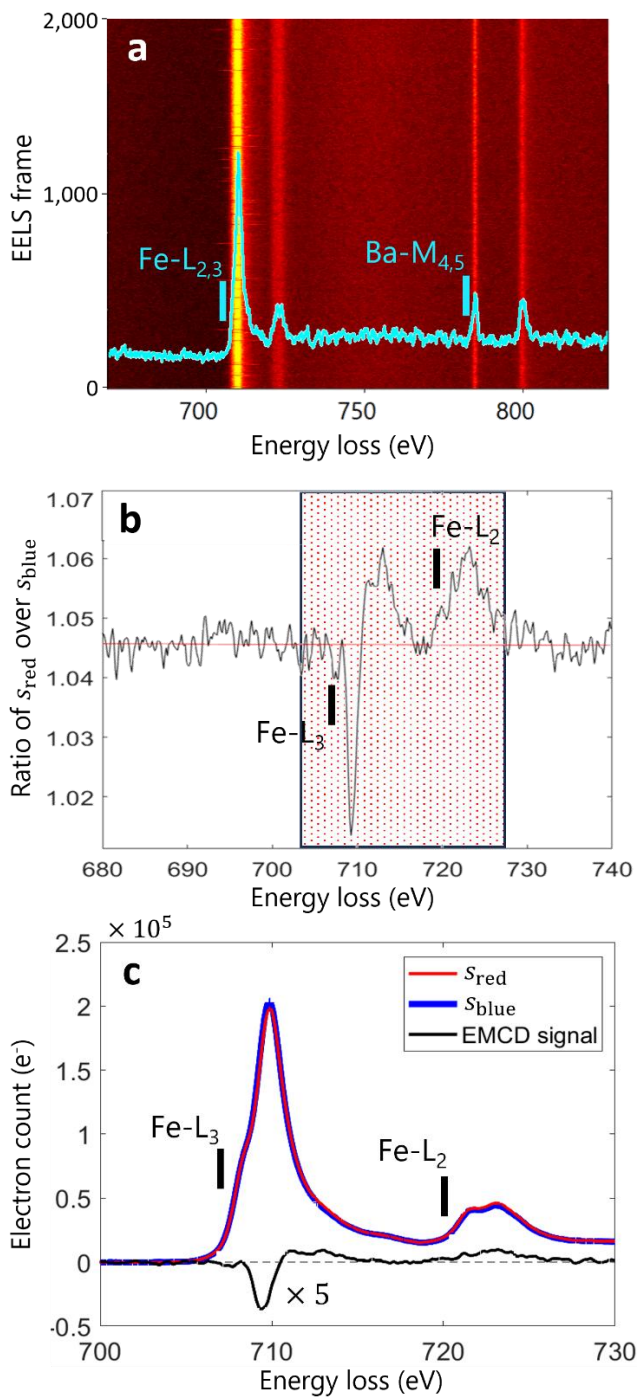

**Figure** 3: Experimental EELS data acquisition and EMCD signal extraction. **(a)** Time series of EELS spectra aligned by cross-correlation. A single spectrum is overlayed as light blue line; the Fe-$L_{2,3}$ and the Ba-$M_{4,5}$ edges are included in this energy region. **(b)** Linear normalization of the ratio of two spectra (black line) using a linear fit function (red line) in the Fe-L pre- and post-edge region. The fitting range excludes the red shadowed part of the spectrum. **(c)** Difference between "red" and "blue" chiral spectra and resulting EMCD signal. EMCD signal enlarged by a factor of 5 (black line) resulting from the difference of the two averaged preprocessed "red" and "blue" chiral spectra.

## Extraction of the EMCD signal

It is crucial to confirm the chiral dichroism for all chiral combinations in order to assess the reliability of the obtained EMCD signal. The EMCD signals obtained by permuting all chiral combinations are shown in Figure 4. A comparison of the top and bottom rows shows that the chiral $(--)$ spectrum plays a significant role in containing the Fe-$L_2$ EMCD component and the shoulder (710 eV to 715 eV) of Fe-$L_3$ stemming from $2p_{1/2}$ and $2p_{3/2}$ ground state, respectively.

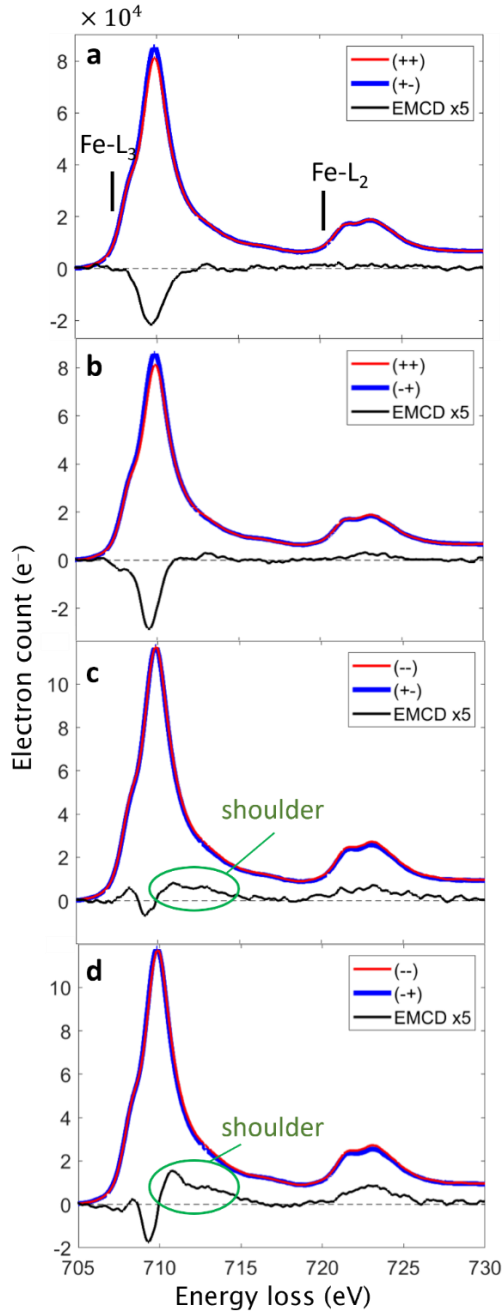

**Figure** 4: All possible EMCD signals resulting from subtractions of opposite chirality EEL spectra. **(a-d)** The EMCD signal (black line), as determined by subtraction of an averaged "red" chiral from an averaged "blue" chiral spectrum respectively, showing a clear EMCD signal including opposite signs at the Fe-$L_3$ and the Fe-$L_2$ edges, respectively (mainly for **c**, **d**).

**Energy misalignment error estimation and its minimization by energy upscaling**

With the chosen energy dispersion of 0.15 eV/channel for the EEL spectra, the energy uncertainty due to sampling is ±0.075 eV. Even such a small energy uncertainty can lead to a critical failure in quantitative EMCD analysis as will be described in this section. Therefore, adequate energy upsampling is required. The amount of upsampling should be determined by whether or not the energy errors exceed a certain threshold. The EMCD signal error resulting from the energy misalignment can be estimated by introducing an artificially energy shift $\Delta E$ between "red" and "blue" chiral spectra. The artificially energy shifted "blue" chiral spectrum denotes $s'_{\text{blue}}(E, \Delta E)$. Thus, the deviation of EMCD signal due to finite energy shift (sampling) can be calculated by subtracting $s'_{\text{blue}}(E, \Delta E)$ from $s_{\text{red}}(E)$. To include the impact of energy uncertainty due to positive and negative energy misalignment, we average the impact of both positive and negative energy shifts:

$$\Delta s'_{\text{EMCD}}(E, |\Delta E|) = \frac{1}{2}(|s_{\text{red}}(E) - s'_{\text{blue}}(E, +\Delta E)| + |s_{\text{red}}(E) - s'_{\text{blue}}(E, -\Delta E)|) \quad (4)$$

By defining signal-to-deviation ratio (SDR) as $s_{\text{EMCD}}(E)/\Delta s'_{\text{EMCD}}(E, |\Delta E|)$, we can evaluate the magnitude of the EMCD signal error due to energy uncertainty. Since this signal-to-deviation ratio drastically changes before and after Fe-$L_3$ peak, we define three unique spectral regions in the EMCD signal as shown in Figure 5 (a). Part 1 (705~710 eV) is defined as pre-Fe-$L_3$ peak region, part 2 (710~715 eV) is defined as post-Fe-$L_3$ peak and part 3 (727~719 eV) includes the whole Fe-$L_2$ peak. We calculate the average of the signal-to-deviation ratio for these three parts as described in Equation 4 and plot them in Figure 7 (b)

$$P_1(|\Delta E|) = \frac{dE \text{ [eV]}}{710 \text{ eV} - 705 \text{ eV}} \sum_{E=705 \text{ eV}}^{710 \text{ eV}} \frac{s_{\text{EMCD}}(E)}{\Delta s'_{\text{EMCD}}(E, |\Delta E|)}$$

$$P_2(|\Delta E|) = \frac{dE \text{ [eV]}}{715 \text{ eV} - 710 \text{ eV}} \sum_{E=710 \text{ eV}}^{715 \text{ eV}} \frac{s_{\text{EMCD}}(E)}{\Delta s'_{\text{EMCD}}(E, |\Delta E|)} \quad (5)$$

$$P_3(|\Delta E|) = \frac{dE \text{ [eV]}}{727 \text{ eV} - 719 \text{ eV}} \sum_{E=719 \text{ eV}}^{727 \text{ eV}} \frac{s_{\text{EMCD}}(E)}{\Delta s'_{\text{EMCD}}(E, |\Delta E|)}$$

where d$E$ indicates the energy dispersion after energy upsampling. As seen in Figure 5 (b), $P_1(|\Delta E|)$ has the lowest SDR, so the lower threshold should be defined for this part. Defining a SDR > 5 as a threshold following a stricter Rose-criteria [29], we should use an energy dispersion of 0.007 eV/channel or smaller. This corresponds to 22 times energy upsampling in our case.

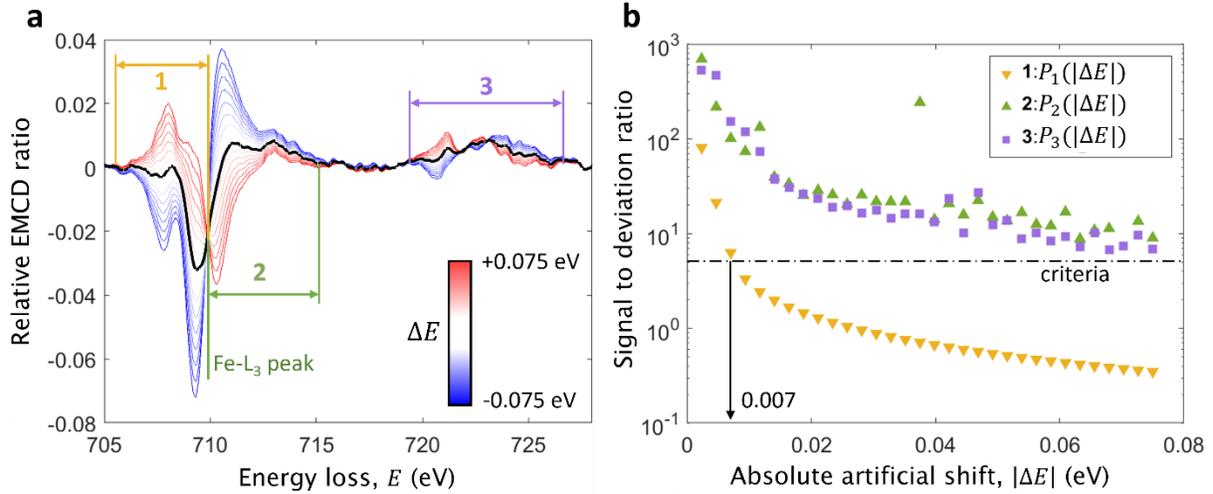

**Figure 5**: Impact of the spectrum shift on the EMCD signal **(a)** Calculated EMCD signals for shifted spectra with a shift ranging from -0.075eV to +0.075eV. The black solid line is the original EMCD signal $i_{EMCD}$ and the red and blue lines show the EMCD signal $i'_{EMCD}$ calculated from shifted spectra for positive and negative shifts, respectively. The energy ranges 1~3 have different shift-impact characteristics thus should be calculated signal-to-deviation ratio (SDR) individually as $P_1(|\Delta E|) \sim P_3(|\Delta E|)$. **(b)** SDR resulting from absolute artificial shift $|\Delta E|$. The dashed horizontal line corresponds to the criteria (SDR > 5), giving a dispersion of 0.007 eV/channel at the intersection at $P_1(0.007) = 5$.

# 3. Noise sources analysis

## 3.1 SNR evaluation

For quantitative measurements, it is essential to know the minimum number of electrons required for a valid EMCD signal. In EELS and EDX studies, elemental identification commonly employs the minimum detection threshold of SNR = 3, also known as the Rose criterion [29,30]. However, this criterion is too weak for the quantitative decomposition of the EMCD signal into several overlapping spectra of different oxidation states. Therefore, we use a stricter threshold of SNR = 5. To evaluate the EMCD SNR, we (i) analyzed the Fe-$L_2$ edge, because the EMCD signal of this edge is weak and often missing, leading to misinterpretations of the EMCD data. And we (ii) evaluated the SNR at two points: at the ascending (E=721 eV) and descending slope (E=725 eV) of the Fe-$L_2$ edge, which correspond to the energies that confine the full-width of half maxima (FWHM) of the EMCD Fe-$L_2$ peak in our study. The standard deviation at both energy channels are statistically derived by the Bootstrapping method [31,32]. This Bootstrap method is a resampling technique that estimates the standard deviation by treating the original dataset as a population and repeatedly drawing samples from it. The process involves creating multiple bootstrap samples (2,000 in our case) of the same size as the original dataset, and calculating the standard deviation for each of these resampled datasets. This can provide a more robust estimate of the standard error of the original dataset than Jack-knife method [33]. Figure 6 shows the SNR evolution with increasing the number of accumulated chirally-scattered electrons as calculated from the accumulated EELS detector count by the gain factor $G_{Faraday}$ measured by a Faraday-cup. To determine the total electron number, at which the SNR = 5 criterion is fulfilled, we fit obtained SNR by a square root function under the assumption that the standard deviation is proportional to the square root of the number of the sample in the Poisson distribution. The fit indicates that 1,035 electrons are required at the ascending and descending points of the Fe-$L_2$ edge to allow a valid quantitative analysis. Using a relative EMCD ratio of 3.55 % (ratio of maximum intensity of the absolute EMCD signal over the maximum intensity of the EEL spectrum averaged for all chiral spectra), this critical minimum electron count corresponds to $2.92 \times 10^4$ electrons in the original chiral EEL spectrum.

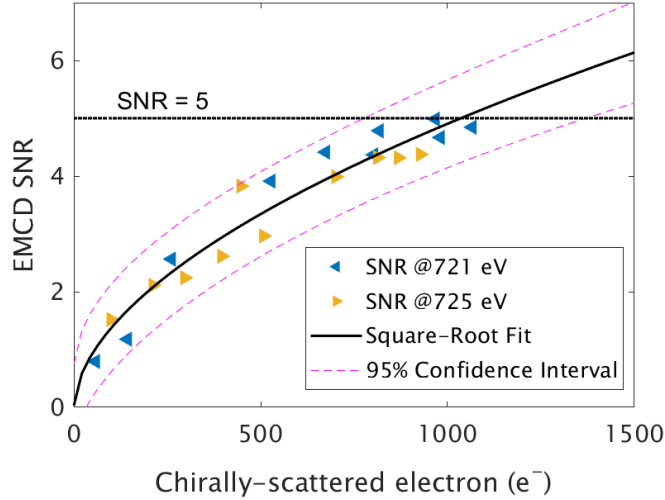

**Figure** 6: SNR evolution by increasing the accumulated chirally-scattered electron number. Each SNR points were calculated from small subsets included in the original dataset by Bootstrapping method. Yellow and blue dots show SNRs at the ascending point (721 eV) and the descending point (724 eV) of the Fe-$L_2$ edge, respectively. Black solid line and magenta dashed lines show the fitted curve and its 95% confidence interval.

## 3.2 Detector noise sources

It is crucial to accurately subtract the EEL spectra from each other because the resulting EMCD signal derived from the EELS spectrum amounts to only a few percent of the original spectrum, making it susceptible to misinterpretations as an artifact. One way to evaluate its accuracy is to assess the noise behavior of the obtained signal. The noise behavior of the EMCD signal is mainly dominated by dark current noise, the shot noise of the chirally scattered primary electrons, and its multiplicative amplification by the detector noise. Dark current noise is caused by thermal electron-hole pair generation in the detector. Detector noise originates from the random generation of photons and electron-hole pairs in the scintillator and the CMOS detector, respectively, as well as the noise of the amplifier integrated into the detector's circuit. Pure shot noise, which originates from the fundamental nature of the quantum detection process, follows spatially uncorrelated Poisson statistics $v = m$, where $v$ indicates the variance of the signal and $m$ indicates the mean intensity of the signal at each position of the detector. The reflection layer, the scintillator, the fiber optics, the active area of the CMOS chip and the readout electronics of the detector, represent additional inevitable sources of noise that modify the shot noise in a cascading manner [34]. We used a simplified multiplicative model of these noise statistics that consists of an effective increase of the noise according to $v = \gamma_{\text{model}} \cdot m$, where $\gamma_{\text{model}}$ denotes the noise amplification coefficient. We discuss this noise model and the derivation of $\gamma_{\text{model}}$ in the Appendix A. By comparing $\gamma_{\text{model}}$ and the noise amplification coefficients measured in the EELS and EMCD experiments, we can estimate the dominant noise source and the quality of the data.

## 3.3 Variance of EMCD signals

According to the laws of error propagation, the difference of "red" and "blue" chiral spectra – the EMCD spectrum – have a variance that is the sum the chiral spectra variances. Thus, this raw variance of EMCD signal, $v_{EMCD}(E)$, that can be estimated by the Bootstrap method, includes not only the subtracted EMCD signal contribution but also the original EEL spectrum contribution. Notwithstanding, the variance of the EMCD signal alone, $v'_{EMCD}(E)$, can be estimated from $v_{EMCD}(E)$ as:

$$v'_{EMCD}(E) = |r_{EMCD}(E)| \cdot v_{EMCD}(E) \qquad 6$$

where $r_{EMCD}(E)$ denote energy-dependent relative EMCD ratio, which is the ratio of the EMCD signal to summed EEL spectrum over each aperture. See the Appendix B to address the full transformation. Using this $v'_{EMCD}(E)$, we can describe the noise of the EMCD signal as:

$$v'_{EMCD}(E) = \gamma_{EMCD} \cdot |m_{EMCD}(E)| \qquad 7$$

where $\gamma_{EMCD}$ is the noise amplification coefficient for the EMCD signal and $|m_{EMCD}(E)|$ is an absolute mean of EMCD signal.

## 3.4 Comparison of the variance-mean plot between flat illuminated image, chiral EEL spectra and EMCD signal

By comparing the variance to mean plots of (a) the flat illuminated image, (b) the EEL spectra from each EELS aperture positions, and (c) the EMCD signal, we can evaluate the main source of uncertainty and the quality of the EMCD experiment. The results are shown in Figure 7 (a)-(c). The variance of the EEL spectra and the EMCD signal are calculated using the Bootstrap method applied to a dataset of 2000 EELS frames as explained in Section 3.1.

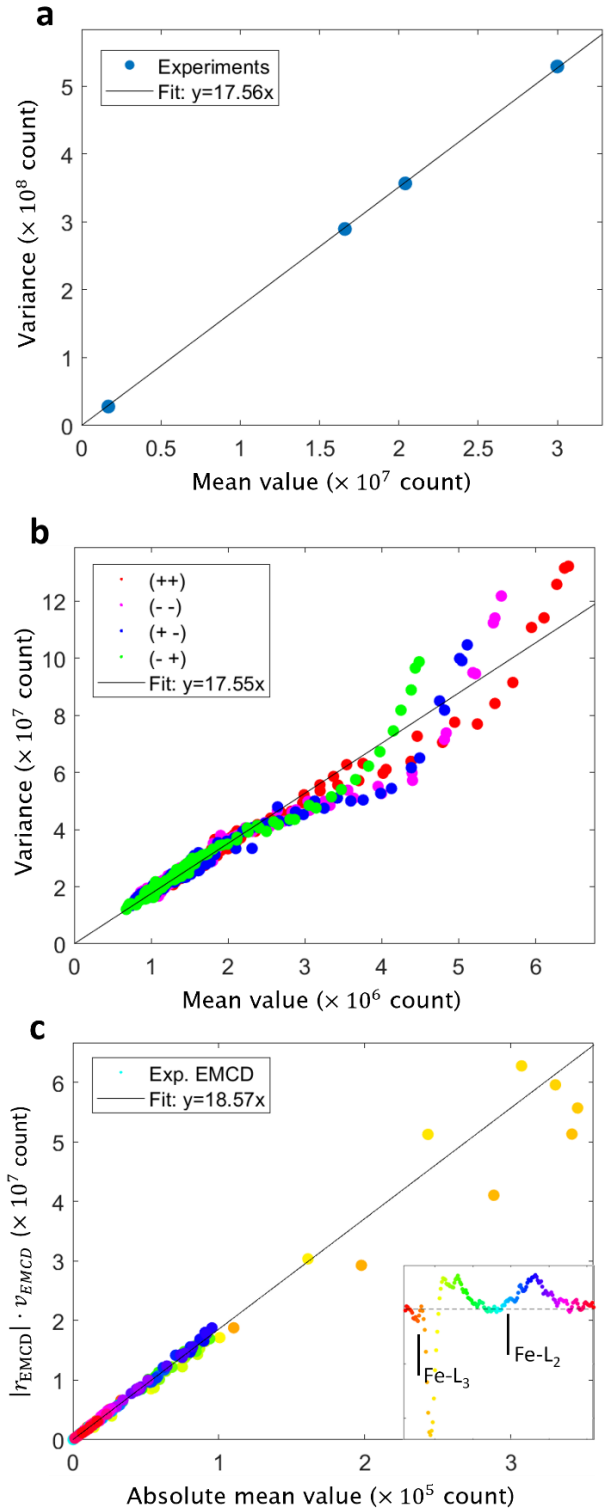

**Figure** 7: **(a)** Variance to mean plot of the flat illuminated image. **(b)** Variance to mean plot of chiral EEL spectra. The line fit was calculated from the energy pixels of all the chiral EEL spectra. **(c)** Variance to mean plot of the EMCD signal. The vertical axis is the estimated variance from EMCD signal as described in Equation 6. The horizontal axis is absolute mean value of EMCD signal. Color coding reflects the energy axis in inset EMCD spectrum.

Table 1: The noise amplification coefficients comparison between the calculation results from the gain $G_{\text{Faraday}}$ by the aforementioned noise model, the flat illuminated image, the chiral EEL spectra from each aperture position, and the EMCD signal.

| $\gamma_{\text{model}}$ | $\gamma_{\text{flat}}$ | $\gamma_{\text{EELS}}$ | $\gamma_{\text{EMCD}}$ |
|---|---|---|---|
| 16.69 | 17.56 | 17.55 | 18.57 |

In the variance to mean plot of the EEL spectra and the EMCD signal, there are slight discrepancies from the linear trend at the higher intensity range, far from flat illumination conditions. To explain this discrepancy, the spatially correlated covariance between the input intensity and the detector noise propagator must be taken into account, which is beyond the scope of this study. The measured noise amplification coefficients from all experiment data are slightly higher than the coefficient from the noise model $\gamma_{\text{model}}$, as shown in Table 1. This indicates the presence of additional uncertainties, such as readout-noise and residuals of the dark-current and gain deviation. Considering the additional uncertainty sources of EELS and EMCD signals, the relationship $\gamma_{\text{Faraday}} < \gamma_{\text{flat}} \approx \gamma_{\text{EELS}} < \gamma_{\text{EMCD}}$ is reasonable. However, $\gamma_{\text{model}}$ which originates from the detector amplified shot noise still shows up as the most dominant source of noise, proving the above made assumptions of the SNR evaluations.

# Summary

We have developed a novel experimental metrology that considers the impact of energy misalignment on the EMCD signal and evaluates noise for the individual energy channels of the detector in classical EMCD measurements, enabling robust quantitative analysis. This methodology was demonstrated using a complex oxide system: Ti-doped barium hexaferrite lamellae, sectioned perpendicular to the (001) crystallographic axis. The SNR of the EMCD signal was quantitatively determined using the bootstrap method applied to the acquired EELS time series. This analysis confirmed that critical EMCD signal features, such as the Fe-$L_{2,3}$ peaks, satisfied the stricter Rose criterion (SNR = 5) for reliable detection. Furthermore, noise propagation in the detector was investigated using a modulated transfer function (MTF) and a noise power spectrum (NPS), revealing that the dominant noise source was detector amplified Poisson noise, thereby demonstrating that the method approaches the statistical limit of measurement precision.

The impact of energy misalignment between the "red" and "blue" chiral spectra—often overlooked in conventional studies—was systematically evaluated. A procedure for estimating and applying the necessary energy upsampling was proposed to mitigate this effect. These improvements ensure reliable quantification of the iron oxidation states within the sample. This metrology can also be applied to other scintillator EELS detectors, not just the Gatan Continuum® Dual EELS/EFTEM camera system used in this study. This can be done by determining the actual gain values, $G_{\text{Faraday}}$, and the detector characteristics, such as the NPS and the MTF. The results of this study remain unchanged in EMCD experiments using direct electron detectors, whose usefulness has recently become widely recognized [35]. Although energy wobbling to avoid noise accumulation caused by gain variations is unnecessary for direct electron detectors due to their electron counting technologies, energy misalignments on the order of meV still occur due to fluctuations in the acceleration voltage and in the diffraction conditions. Therefore, upsampling remains an important alignment procedure.

In summary, accurately identifying the various sources of error in quantitative EMCD measurements is crucial for reliably characterizing new magnetic materials, especially those with unknown magnetic structures. Conventional EMCD studies have relied on the $m_L/m_S$ ratio to validate the correctness of the EMCD signal. However, this approach is not applicable to new magnetic materials with unknown $m_L/m_S$ ratios or to complex oxides with overlapping atomic sites. Furthermore, evaluating the entire EMCD energy range based on a single representative scalar value can be misleading. This study offers an alternative evaluation perspective based purely on statistical analysis.

# Appendix A: Noise propagation of the CMOS detector

In an ideal detector, an impinging electron is recorded as a delta-like function (sampled by discrete pixels). Thus, the fine structure of the signal (i.e., high-frequency components) is preserved because the Fourier transform of a delta function is a constant. However, an actual stochastic detector has several input signal propagation paths that blur the fine structure of the signal. To evaluate this blurring process, we have to measure the detector's signal transfer function, in other words, a ratio of the output signal to the input signal in frequency space is called the modulated transfer function $\mathrm{MTF}(k_x, k_y)$. The zero frequency value of the MTF(0,0) corresponds to the true gain $G_{\mathrm{Faraday}}$ [36]. The gain of the CMOS detector used in this work was measured with the help of a Faraday cup, yielding $G_{\mathrm{Faraday}} = 51.7$.

In our case, we use the edge method to determine the MTF [36]. Here, the beam stopper above the GIF camera blocks part of the beam to visualize the electron spreading in the scintillator Therefore, only the 1D line spread function (LSF) can be calculated, as shown in Figure A1 (a). To prevent the stopper shadow from aliasing onto the detector pixels, the knife edge was slightly inclined against the pixel rows. Deviations of the LSF from the original PSF are caused by Fraunhofer diffraction and pixel integration in *y*-direction, which could be reduced by using single-spot illumination method introduced by Niermann and Lubk [34,37]. The obtained PSF is blurred from the original PSF, and the MTF is reduced in the high frequency range due to the sampling effect.

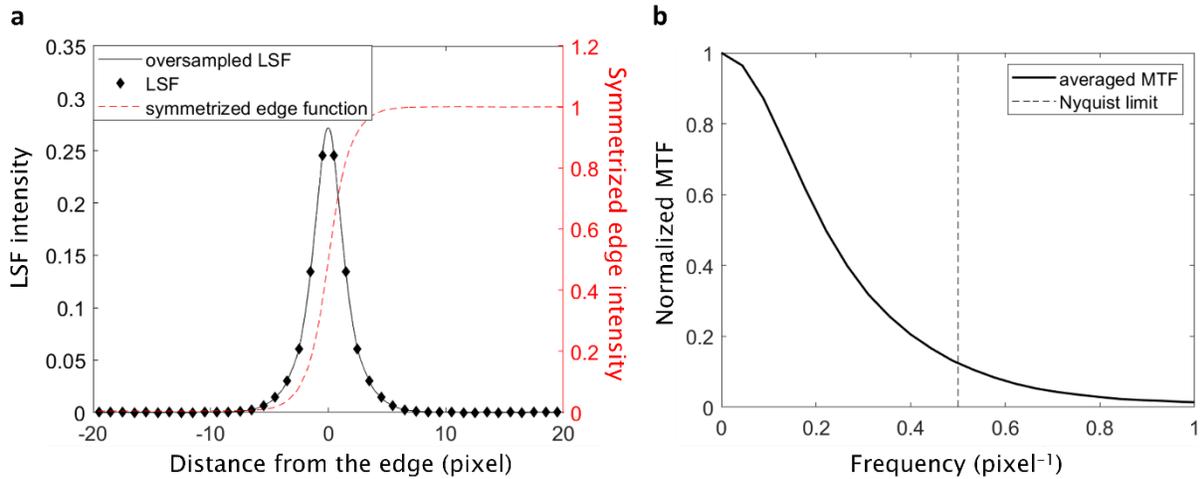

**Figure** A1: **(a)** Oversampled symmetrized edge function (red bleak line) and an obtained LSF (black line with diamonds). This obtained line profile of the edge shows an almost perfect point symmetry. However, the Poisson noise is stringer on the bright side than on the dark side. Hence, the edge profile was symmetrized with weighting

factor which is inversion of standard deviation of each pixels [36]. **(b)** Oversampled MTF of our detector at 200 kV acceleration voltage. Slightly canted beam stop allows oversampling beyond Nyquist limit (= 0.5 /pixel$^{-1}$).

To consider the noise transfer behavior in the detector, we start with the simplest situation. In an ideal detector, a flat image exhibits only spatially uncorrelated white noise that follows a Poisson distribution. Therefore, the calculated variance $v$ is equal to the mean value of the image $m$, if there is no gain. In deterministic detector, where each incoming electron is amplified in exactly the same way, the mean and the standard deviation are increased by the gain $G_{\text{Faraday}}$. Thus, the variance is obtained by multiplying the mean value by the gain as:

$$v = G_{\text{Faraday}} \cdot m \qquad \text{A.1}$$

In a stochastic detector with a PSF, neighboring pixels are (positively) correlated. This correlation reduces the variance compared to the mean value by the correlation factor $C_{\text{corr}}$ [34,38]. The correlation factor, $C_{\text{corr}}$, can be calculated using noise power spectrum (NPS) analysis. Two flat illuminated images with the same exposure are subtracted from each other to obtain a noise image with twice the variation of the original flat illuminated image without artifacts from the gain correction or dark correction. The NPS can be calculated by taking the square of the Fourier transformation of the obtained noise image [38]. The obtained NPS for different mean intensities (shown in Figure A2 (a)) exhibit a reduction in the high frequency region due to PSF correlation. Thus, a correlation factor can be obtained by integrating up to the Nyquist limit and dividing the ideal NPS (constant) and the actual NPS as:

$$C_{\text{corr}} = \frac{\sum_{k_x,k_y} 1}{\sum_{k_x,k_y} \text{NPS}(k_x, k_y)} = 3.10 \qquad \text{A.2}$$

$C_{\text{corr}}$ is reduced when the neighboring PSFs are not saturated as shown in Figure A2 (b). When there is no signal, only uncorrelated dark current noise remains ($C_{\text{corr}}(m = 0) = 1$). This reduction in correlation can be problematic when analyzing the noise properties of ultra-low count EEL spectra, such as EMCD measurements. The EELS exposure time is increased to 0.5 s, where the acquired chiral EEL spectra has sufficient intensity to saturate $C_{\text{corr}}$ as shown in Figure A2 (b). Consequently, the constant value $C_{\text{corr}} = 3.10$ can be used in later analyses.

Thus, the modeled noise amplification coefficient can be determined as $\gamma_{\text{model}} = G_{\text{Faraday}}/C_{\text{corr}} = 16.69$ in this detector.

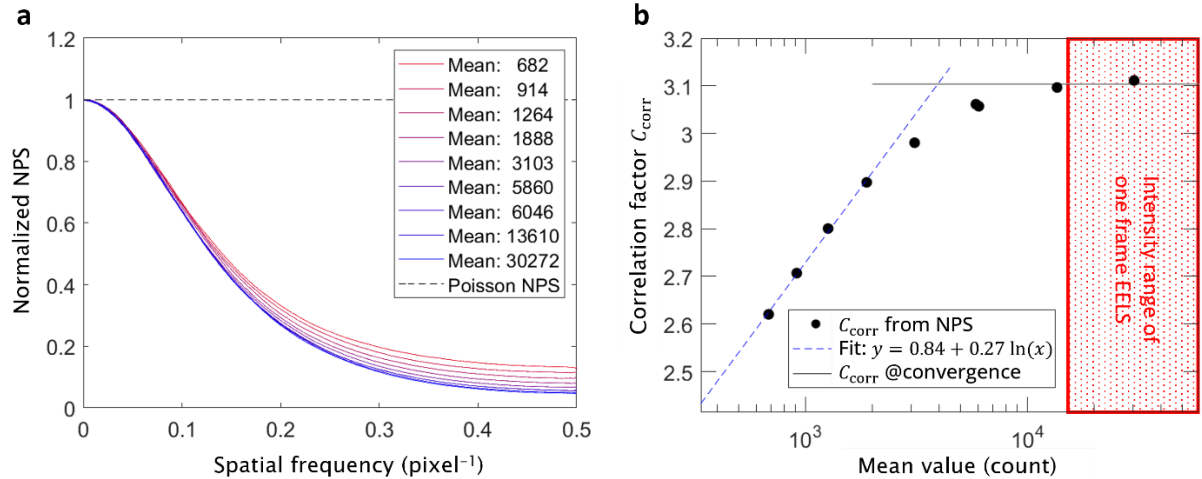

**Figure** A2: **(a)** NPS for different mean intensities and NPS for the shot noise (constant). The discrepancy in high frequency results in the correlation factor $C_{\text{corr}}$ to correct the variance. **(b)** Semi-log plot for the correlation factor $C_{corr}$ determination for varying intensities. Intensity range of one frame of a chiral EEL spectra dataset is shown as a red box.

# Appendix B: Variance estimation of EMCD signal

The mean EMCD signal $m_{\mathrm{EMCD}}(E)$ is calculated as

$$m_{\mathrm{EMCD}}(E) = m_{++}(E) + m_{--}(E) - m_{+-}(E) - m_{-+}(E) \tag{B.1}$$

where the $m_{++}(E)$, $m_{--}(E)$, $m_{+-}(E)$, and $m_{-+}(E)$ denote means of spectra from each EELS entrance aperture position. According to the error propagation formula, the raw variance of the EMCD signal $v_{\mathrm{EMCD}}(E)$ at energy position $E$ is described as:

$$\begin{aligned} v_{EMCD}(E) &= \left(\frac{\partial m_{EMCD}(E)}{\partial m_{++}(E)}\right)^2 v_{++}(\varepsilon) + \left(\frac{\partial m_{EMCD}(E)}{\partial m_{--}(E)}\right)^2 v_{--}(\varepsilon) \\ &+ \left(\frac{\partial m_{EMCD}(E)}{\partial m_{+-}(E)}\right)^2 v_{+-}(E) + \left(\frac{\partial m_{EMCD}(E)}{\partial m_{-+}(E)}\right)^2 v_{-+}(E) \end{aligned} \tag{B.2}$$

All squared partial derivative coefficients become 1 from Equation (B.1). Variances of each aperture position can be substituted by mean spectra using the noise model

$$\begin{aligned} v_{\mathrm{EMCD}}(E) &= v_{++}(E) + v_{--}(E) + v_{+-}(E) + v_{-+}(E) \\ &= \gamma \cdot (m_{++}(E) + m_{--}(E) + m_{+-}(E) + m_{-+}(E)) \end{aligned} \tag{B.3}$$

By multiplying both sides of Equation (B.3) by an absolute mean of EMCD signal $|m_{\mathrm{EMCD}}(\varepsilon)|$, we obtain the first row of Equation (B.4). Since sum of each chiral spectra is bigger than zero, $(m_{++}(\varepsilon) + m_{--}(\varepsilon) + m_{+-}(\varepsilon) + m_{-+}(\varepsilon))$ can be divided on both sides of the equation:

$$\begin{aligned} v_{EMCD}(E) \cdot |m_{\mathrm{EMCD}}(E)| &= \gamma \cdot (m_{++}(E) + m_{--}(E) + m_{+-}(E) + m_{-+}(E)) \cdot |m_{\mathrm{EMCD}}(E)| \\ v_{\mathrm{EMCD}}(E) \cdot \frac{|m_{\mathrm{EMCD}}(E)|}{m_{++}(E) + m_{--}(E) + m_{+-}(E) + m_{-+}(E)} &= \gamma \cdot |m_{\mathrm{EMCD}}(E)| \end{aligned} \tag{B.4}$$

Here, we defined energy-dependent absolute relative EMCD ratio $|r_{\mathrm{EMCD}}(E)|$ for easy understanding. Thus, the variance of EMCD signal alone was estimated as:

$$|r_{\text{EMCD}}(E)| \equiv \frac{|m_{\text{EMCD}}(E)|}{m_{++}(E) + m_{--}(E) + m_{+-}(E) + m_{-+}(E)}$$

$$|r_{\text{EMCD}}(E)| \cdot v_{\text{EMCD}} = \gamma \cdot |m_{\text{EMCD}}(E)|$$

$$v'_{\text{EMCD}}(E) = |r_{\text{EMCD}}(E)| \cdot v_{\text{EMCD}}$$

B.5

# Acknowledgement

This work is funded by the Deutsche Forschungsgemeinschaft (DFG, German Research Foundation) grant no. 504660779, through TRR 404 Active-3D (project number 528378584) and the priority program SPP 2137 (project number 403503416). We thank Alexander Tahn for support with sample preparation. Axel Lubk thanks for funding through CRC1143 (project number 247310070). We are also grateful to V. A. M. Brabers for providing the $BaFe_{11}TiO_{19}$ single crystal.